\documentclass{aa}    
\usepackage{graphics}

\begin{document}

\thesaurus{05 ( 10.15.01; 05.01.01; 08.22.01)}

\title{Absolute Proper Motions of Open Clusters.}
\subtitle{I. Observational data}

\author{H.\ Baumgardt\inst{1} \and C.\ Dettbarn\inst{2} \and R.\ Wielen\inst{2}}
\offprints{H.\ Baumgardt}
\institute{Department of Mathematics and Statistics, University of Edinburgh,
Edinburgh EH9 3JZ, UK
\and
Astronomisches Rechen-Institut Heidelberg, M\"onchhofstra{\ss}e
 12-14, D-69120 Heidelberg, Germany\\
e-mail: holger@maths.ed.ac.uk}

\date{Received / Accepted }

\maketitle

\begin{abstract}
Mean proper motions and parallaxes of 205 open clusters were determined from their member stars found in  
the Hipparcos Catalogue. 360 clusters were searched for possible members, 
excluding nearby clusters with distances $D < 200$ pc. Members were selected using
ground based information (photometry, radial velocity, proper motion, distance from 
the cluster centre) and information provided by Hipparcos (proper motion, parallax).
Altogether 630 certain and 100 possible members were found. 
A comparison of the Hipparcos parallaxes with photometric distances 
of open clusters shows good agreement. The Hipparcos data 
confirm or reject the membership of several Cepheids in the studied clusters.

\end{abstract}

\section{Introduction}

The aim of this work is to determine proper motions and parallaxes of open clusters
using the Hipparcos Catalogue (ESA \cite{esa}). We will focus here on the more
distant open clusters and try to find proper motions for all clusters with distances
greater then $D=200$ pc.
In combination with accurate photometric distances and radial velocities, 
the proper motions will provide valuable information on 
the kinematic parameters of the galactic rotation curve and the distance to the galactic centre.
They are also useful to study the formation and evolution of the open cluster system.
These questions will be addressed in a forthcoming paper. The present paper 
provides the observational database for these studies.

Stars in open clusters were searched and proposed by Wielen \& Dettbarn (\cite{wd85}) and
selected by a specific working group within the INCA Consortium during the creation
of the Hipparcos Input Catalogue. However, the Hipparcos Catalogue
may contain additional cluster stars not selected by this working group: First, 
there may exist bright cluster stars that have so far not been studied at all or 
were studied in detail only 
after the creation of the Hipparcos Input Catalogue. Such stars could have entered the
Hipparcos Catalogue as part of the all sky survey.
In addition, cluster stars may have been proposed by other research groups if they are 
of astrophysical interest. 

It therefore seemed appropriate to perform a new search for stars in open 
clusters. Sections 2 and 3 describe the selection of the open clusters
and the search for Hipparcos stars in the cluster fields. 

Section 4 describes the way in which cluster members were identified among the candidates
and the next section describes how the mean astrometric parameters of the clusters
were determined.
Section 6 lists the classification of the Hipparcos stars 
and presents the mean astrometric parameters of the open clusters. 
Two applications of the astrometric parameters are also described in this
section: First, we compare the Hipparcos parallaxes with the photometric parallaxes of
the open clusters to check for systematic errors in both methods.
Second we use the proper motions of the open clusters to check the 
membership of several Cepheids. Section 7 summarises our results.

\section{Cluster selection}

The basic data of the open clusters (positions, distances, magnitudes of
brightest stars) were taken from the catalogue of Lyng{\aa} (\cite{lyn87}).
1151 open clusters are listed in his catalogue, most of which however are too
faint for Hipparcos. We therefore removed
all clusters that have never been studied at all, since they are presumably very faint
and even if they contain stars bright enough for Hipparcos, there is no information
available to identify them as members.
In addition, we omitted clusters where, according to Lyng{\aa} (\cite{lyn87}),
brightest members are fainter than $V = 12$ mag,
since it is very unlikely that they contain Hipparcos stars.
Nearby clusters with distances $D < 200$ pc were also
omitted from our analysis. This was done because they have 
large angular diameters,
so that the assumption of a common proper motion of all cluster stars
is not valid for them. They require a different kind of analysis
(for example convergent point methods like in Perryman et al.\ 
\cite{pman}), which is beyond the scope of the present paper. Nearby 
clusters not discussed here include the Hyades, Pleiades, Coma Ber,
IC 2391, IC 2602, $\alpha$ Persei Cluster, Praesepe (NGC 2632) and the 
UMA Star Cluster. Hipparcos results for most of them can already be found in the literature
(Perryman et al.\ \cite{pman}, van Leeuwen \cite{fvl}, Robichon et al.\ \cite{rob}). 
Our final list contained 360 clusters which may have members in the
Hipparcos Catalogue.

We note that Lyng{\aa} (\cite{lyn87}) lists two additional clusters with distances
less than 200 pc (Col 399, Upg 1), but Baumgardt (\cite{baumg})
has recently shown on the basis of proper motions from the Hipparcos
and ACT (Urban et al.\ \cite{urban}) catalogues that they do not exist at all.
We finally note that Platais et al.\ (\cite{pla98}) found several new cluster candidates
in the Hipparcos Catalogue, but most of them require further study to confirm
their reality.

\section{Member search}

Stars can only be gravitationally bound to a star cluster if they are 
located inside its tidal radius. If a star cluster is seen
from outside, this corresponds to a spherical region on the sky inside 
which stars must lie in order to be bound to the cluster.
To find the members,
we therefore took a rough estimate for the tidal radius 
(we assumed 12 pc, the tidal radius for a 1000~$\mbox{M}_{\odot}$ star 
cluster in the solar neighbourhood),
the cluster distance from Lyng{\aa} (\cite{lyn87}), and calculated the angular 
diameters of these regions.
They were then searched for Hipparcos stars and 2900 stars were found
altogether.

\section{Member selection}

The crucial part in the determination of the mean astrometric parameters of star 
clusters is the proper separation of members and non-members. This is
especially important in the present case since we will typically have
only a few stars per cluster (sometimes only one star), so that 
a single misclassified field star can already influence our final 
solution considerably. Since the information provided by Hipparcos is
not sufficient to separate the cluster members from the field stars,
we had to judge the membership by combining Hipparcos data with
information provided by ground based studies. These include:

\begin{itemize}
\item Photometry and spectroscopy
\item Proper motions
\item Radial velocities
\item Parallaxes
\item Angular distances from the cluster centres
\end{itemize}

The following paragraphs will illustrate this approach.

\subsection{Photometry and spectroscopy}

Multicolour photometry and spectroscopy were our main criteria for the 
membership determination due to the fact that they are available for most stars
and rule out membership for many of them. We performed a literature
search for each cluster and noted the classification of our candidates
in the various studies. We did not examine the membership on our own, 
since in most photometric studies the data was already carefully 
analysed by the authors themselves.

\subsection{Proper motions}

The main proper motion source is the Hipparcos Catalogue itself.
If the proper motion of a cluster is already known from a few
certain members in the Hipparcos Catalogue, it is possible to check the 
membership of the remaining stars with the proper motion of the cluster: New members are expected to 
have proper motions in agreement with the known members. However, the majority of our
clusters do not have proper 
motions which clearly separate them from the field stars, therefore a 
matching proper motion in the Hipparcos Catalogue is a necessary but not a sufficient 
criterion for cluster membership.

In some cases we combined the Hipparcos proper motions with proper motions
of additional members found in the TRC Catalogue (H\o g et al.\ \cite{hoeg}). This was
done for clusters where we could find no common motion among
the Hipparcos stars or where only one member could be found in Hipparcos
and we regarded it necessary to check its membership further. We also checked
our proper motions against proper motions derived by Glushkova et al.\ 
(\cite{gl99}) from the TRC Catalogue.

Ground based proper motions were also used for the member determination. They 
are available for about 40 of the studied clusters. Ground based proper motions have 
generally the same or an even higher accuracy than the Hipparcos proper
motions. Unfortunately, since they are not on the Hipparcos system, they cannot be
directly compared with Hipparcos. Due to their accuracy, they were
nevertheless a powerful tool to eliminate field stars from our sample.

\subsection{Radial velocities}

High precision radial velocities are best suited to distinguish between members 
and non-members. Since they are available for relatively few stars only, 
their application is limited to a small number of clusters.
They nevertheless give valuable information, since the
studied stars are often giants or variable stars for which
photometry and spectroscopy are of limited use.

\subsection{Parallaxes}

Parallaxes are provided by the Hipparcos satellite. They put tight
limits to the distances of nearby stars, but get less and less accurate
the further the star is away. For most clusters they were
able to eliminate a few foreground stars, but the majority of the field stars
have parallaxes which would be compatible with the assumption of a cluster membership.

\subsection{Angular distance from the cluster centre}

While the surface density of a star cluster drops from the inner to the
outer parts, the density of the background remains essentially 
constant. Thus, a star seen close to the cluster
centre is more likely a member than a star near the tidal radius. 
The angular distance therefore gives some hints for the classification
of stars. 
\vspace*{0.3cm}

Our final classification was made by taking into account all the information we could get.   
Stars were divided into three
categories (members, possible members and non-members) according on how well they fulfilled
the membership criteria.

\section{Determination of the mean astrometric parameters}

We neglected perspective effects and assumed that all cluster stars have similar proper 
motions and parallaxes. This is justified by the large distances and resulting 
small angular diameters of the clusters studied. Due to the large distances the internal 
motions of the cluster stars can also be neglected. 

Stars included in the Double and
Multiple Systems Annex of the Hipparcos Catalogue, i.e. with C,G,V,O or X entries in field
H59, were not used to derive the mean astrometric parameters, since the astrometric solution
derived by Hipparcos is affected by the binary nature of these stars, as described in Robichon 
et al.\ (\cite{rob}).

The mean astrometric parameters were kindly calculated for us by Floor van Leeuwen. 
For clusters containing more than one star,
he used the method described in van Leeuwen \& Evans (\cite{vanl}) to calculate the mean
astrometric parameters.
This method takes into account the correlations that exist between
the abscissae residuals of Hipparcos stars measured on the same reference great circles.
These correlations increase with decreasing angular separation and are important
in our case due to the small angular diameters of the clusters studied.
No such correlations exist for clusters containing only one star, so
we took the solution for these clusters directly from the solution given for the 
single member in the Hipparcos Catalogue.

Alternatively, we also calculated the astrometric parameters without taking small-scale 
correlations into account. This was done to estimate the influence of these correlations.
For that purpose, the cluster solutions were obtained from the member stars by taking the 
mean of their Hipparcos solutions.

\section{Results}

For 140 clusters no members could be found in the Hipparcos Catalogue. In the 
remaining 205 clusters,
we found altogether 630 certain and 130 possible members. They are listed in Table 1 together with some 
non-members. The majority of non-members were never supposed to be cluster
stars since they have colours, magnitudes and angular distances that clearly rule out 
their membership. In order to keep Table~1 short,
we therefore list only non-members which could be found in the 'Database
for Galactic Open Clusters' (BDA) by Mermilliod (\cite{mer95}). These stars are 
presumably much closer to the cluster centre than the majority of non-members and many of 
them were previously thought to be members. 

Column 8 of Table 1 gives a membership
probability derived from the proper motions. It was calculated as follows:
If a star was not used to calculate the mean cluster motion, we took  
the difference
\vec{x} between the proper motion of the cluster (taken from Table 2) and the star and calculated
the product of this difference with the sum $\Sigma$ of the covariance matrices of the star 
and the cluster according to:
\begin{equation}
c = x' \; \Sigma^{-1} x \;\;. 
\end{equation}

The dimensionless number $c$ was then transformed into a membership probability under the assumption
that it is distributed like a chi-square distribution with two degrees of freedom. 
If a star was used for the
derivation of the mean cluster motion, we first calculated a solution for the cluster
without using the star in question and compared the proper motion of the star with 
this new solution. 

Most non-members were classified as such due to their large proper motion differences
relative to the mean cluster motion.
The remaining
stars that were classified as non-members were either too far away from the cluster 
centre to be 
bound or had a photometry which
was incompatible with the assumption of cluster membership. Possible members with a high
probability for membership are often not well-studied, outlying stars. They may be members according 
to their proper motion and photometry. 

Table 2 presents our final solution for the astrometric cluster parameters, determined as
outlined in section~5. The unit weight standard deviations show some scatter around the
theoretical value of one. This is caused by the small number of stars available
for most clusters. The mean over all clusters that contain at least two stars is 1.01, 
which is very close to the theoretical value. We note that
the unit weight standard deviations are generally smaller than one for clusters 
with more than 5 stars in the astrometric parameter determination.
Since the correlations of the abscissae residuals become more important for clusters
containing many stars, this may indicate that these correlations were slightly overestimated
when the mean values were calculated. If the small-scale correlations are not taken
into account, we obtain a mean unit weight standard deviation of 0.97 for all clusters,  
which is also very close to the theoretical value of one. 
\begin{figure}
 \resizebox{8cm}{!}{\includegraphics{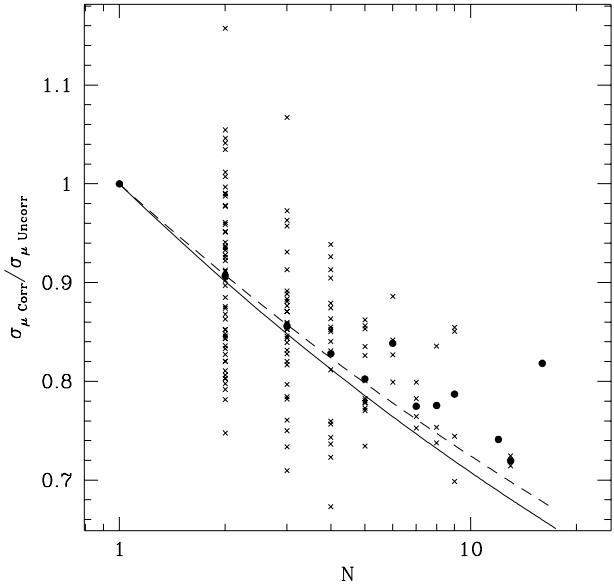}}
 \caption{Proper motion errors derived by taking small-scale correlations into account
compared to the uncorrelated solution as a function of the number of cluster stars $N$. The crosses  
show individual clusters, the dots mark the mean values for a given $N$. The solid line
shows a dependence proportional to $N^{-0.15}$ (exponent derived from Lindegren \cite{li88}), the dashed 
line shows one proportional to $N^{-0.14}$.} 
\end{figure}

Figure 1 compares the errors
obtained by taking small-scale correlations into account with the errors of the uncorrelated 
solution. Shown is the ratio of the sums $\sigma_\mu$ of the 
proper motion errors $\sigma_\mu = \sqrt{\sigma^2_{\mu \alpha*} + 
\sigma^2_{\mu \delta}}$ as a function of the number of cluster stars $N$.  
If a cluster contains only one star, this ratio must be unity, since both solutions
are taken directly from Hipparcos. For clusters with more than one star, the mean error
should be proportional to $N^{-0.5}$ in the uncorrelated case. For the
correlated case, Lindegren (\cite{li88}) estimated an $N^{-0.35}$ dependence, so the
ratio of the errors should be proportional to $N^{-0.15}$. Clusters
containing less than 6 Hipparcos stars are best fitted by a $N^{-0.14}$ law, very
close to the expected value. An $N^{-0.36}$ decrease can therefore be taken as a rule of 
thumb to estimate the errors of the astrometric parameters for correlated measurements in
Hipparcos.

Robichon et al.\ (\cite{rob})
determined absolute proper motions and parallaxes of all clusters that are closer than 
300 pc or have more than 4 members in the Hipparcos Catalogue. Figure 2 compares our 
proper motions with theirs for the clusters in common. Although they also use 
the Hipparcos Catalogue and a similar method to derive the proper motions, the final results
differ by typically 0.5 mas/yr, which is of the same order as the quoted errors. The 
differences 
are due to differences in the stars selected as cluster members and 
slight differences in the abscissae formal errors and correlations
that are used in the methods of Robichon et al.\ (\cite{rob}) and van 
Leeuwen \& Evans (\cite{vanl}).  
   
\begin{figure}
 \resizebox{8cm}{!}{\includegraphics{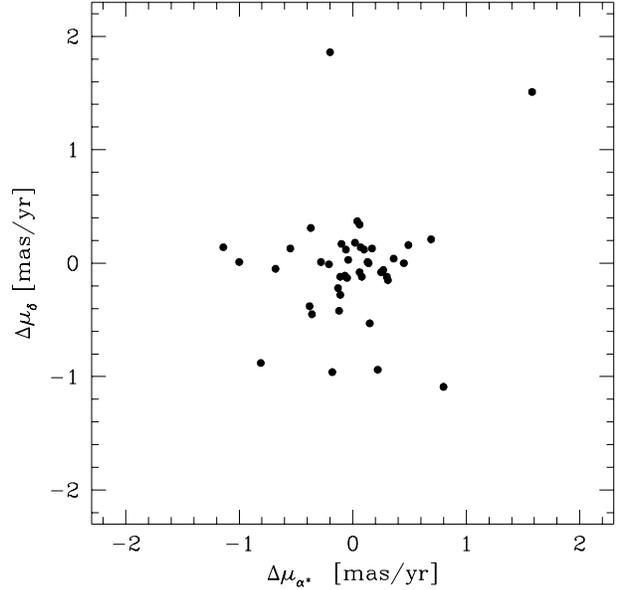}}
 \caption{Plot of the proper motions derived in this work minus the proper motions from 
  Robichon et al.\ (\cite{rob}) for the clusters in common.} 
\end{figure}

\subsection{Comparison of the Hipparcos parallaxes with photometric distances}

Loktin et al.\ (\cite{lm94}, \cite{lok97}) and Dambis (\cite{dam99}) have determined distances
and ages of open clusters on the basis of published photometry. 
Their data represent large and homogeneous parameter sets. With
the help of the Hipparcos parallaxes, we can check for global errors $f$ in their distance
scales:
\begin{equation}
R_{Phot} = f \cdot R_{True}\;\;. 
\end{equation}
Such errors can be detected with a test similar to that used by Feast \& Catchpole
(\cite{feca}): 
The (true) distance of a cluster is connected with the cluster parallax $\pi$ through the 
equation
\begin{equation}
\pi = 1/R_{True} \;,
\end{equation}
so that an estimate for $f$ can be obtained by inserting equation 3 into 2:
\begin{equation}
f = 0.001 \; \pi_{Hip} \; R_{Phot} \;.
\end{equation}
Here the Hipparcos parallax $\pi_{Hip}$ is measured in mas and the photometric distance is in parsecs. 
The mean over all clusters was taken to derive $f$:
\begin{equation}
<\!\!f\!\!> = \frac{\sum \frac{f_i}{\sigma^2_i}}{\sum \frac{1}{\sigma^2_i}}
\end{equation}
The errors $\sigma_i$ on the right-hand side are a combination of the errors in the
Hipparcos parallaxes and the (random) errors in the photometric distances to individual clusters,
taken to be 10\% for the Loktin sample.
Since the error estimate of the photometric distance is done with the observed distance 
and not with the true one, the above
test leads to a slightly biased estimate for $f$. Monte-Carlo simulations show that
this bias remains small as long as the errors in the photometric distances are not of the order 
of 40\% or larger. 

From 186 clusters in common between Loktin et al.\ (\cite{lok97}) and this work, we
derive a correction of $f = 1.12 \pm 0.05$ to their distance scale, i.e. their distances
should be
decreased by 12 \%. Part of this decrease may be explained by the fact that
Loktin et al.\ assumed an Hyades distance modulus of $(M-m)_0 = 3.42$, which is too large
since the Hipparcos 
data indicate $(M-m)_0 = 3.33$ (Perryman et al.\ \cite{pman}). 
The errors in the photometric distances of Dambis (\cite{dam99}) are taken from their work.
From 117 clusters in common, we obtain a correction
factor of $f = 0.99 \pm 0.06$. Their overall distance scale is therefore in very good 
agreement with Hipparcos. 

Using the test of Arenou \& Luri (\cite{alu}), we can also check the  
parallax errors in the Hipparcos Catalogue: the
difference $\Delta \pi$ between the Hipparcos parallax and the photometric parallax 
$\pi_{Phot} = 1/R_{Phot}$
\begin{equation}
\Delta \pi = \pi_{Hip} - \pi_{Phot}
\end{equation}
is mainly caused by the error in the Hipparcos parallax, since the errors in the photometric 
parallaxes are only of the order of 0.1 mas. If the errors in the Hipparcos Catalogue are 
normally distributed
and show no correlations other than those that were already accounted for in section 5,
the ratio of $\Delta \pi/\sigma_{Hip}$ should also be normally distributed.

Figure 3 shows the distribution of $\Delta \pi/\sigma_{Hip}$ for the Loktin et al.\ distances.
A normal distribution provides a very good fit to 
the data. Narayanan \& Gould (\cite{nago}) proposed correlated errors extending over 
angular scales of 2 to 3 degrees and with amplitudes of up to 2 mas in the Hipparcos 
parallaxes as the reason for the discrepancy between recent photometric distances and 
the Hipparcos distance to the
Pleiades. Our clusters have small angular sizes and would be effected by such
errors as a whole. If they exist in the entire Hipparcos Catalogue, such errors would
significantly broaden the cluster distribution in Figure 3 (note that typical
parallax errors of our clusters are only 0.5 mas). Since such a broadening
is not observed we can rule out correlated errors with amplitudes of more than a few tenths
of a mas for the vast majority of our clusters. A similar conclusion was drawn by
Arenou \& Luri (\cite{alu}). We confirm their results with a larger database. 
We finally note that a similar result is obtained if the distances of Dambis (\cite{dam99}) 
are used.
\begin{figure}
 \resizebox{8cm}{!}{\includegraphics{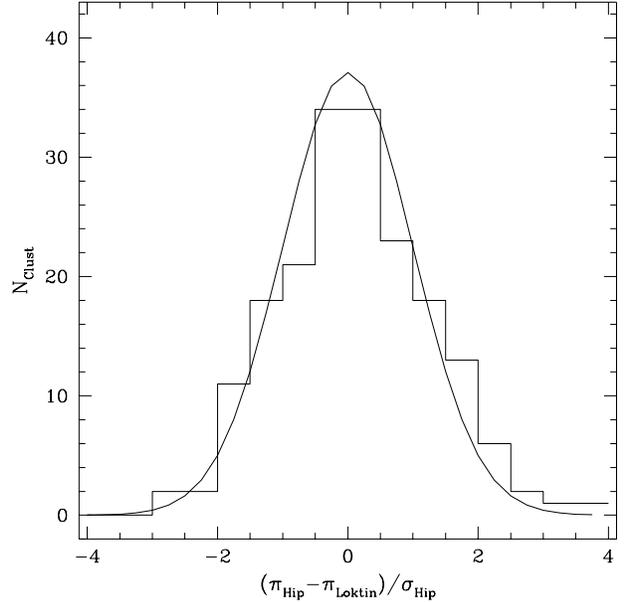}}
 \caption{Histogram of the normalised differences $(\pi_{Hip}-$ $\pi_{Loktin})/\sigma_{Hip}$. A gaussian
 provides a very good fit to the data. Correlated errors of the order of
 0.5 mas or larger would significantly broaden the observed distribution and can therefore be ruled out.}
\end{figure}

\subsection{Cepheids in Open Clusters}

The Hipparcos proper motions (and to a lesser
degree also the parallaxes)
confirm or reject the cluster membership of stars which are of astrophysical interest, like
e.g.\ Wolf-Rayet stars, red giants and various types of variable stars. This may help to better define 
their physical parameters. As an example, we discuss the membership of Cepheids in our clusters.

11 Cepheids are included in Table 1. Many of them are the only Hipparcos star in their 
cluster, so our classification is entirely based on results found in the literature. 
Eight Cepheids are in clusters with two or more members (see Table 3 and Figure 4).

The membership of U Sgr (HIP 90836) in IC 4725 and of S Nor (HIP 79932) in NGC 6087 was already 
discussed by Lyng{\aa} and Lindegren (\cite{ll}) on the basis of the Hipparcos data. We 
confirm the cluster membership of both Cepheids. DL Cas (HIP 2347) is a highly probable member of 
NGC 129 on the basis of its photometry (Turner et al.\ \cite{tfp92}), radial velocity 
(Mermilliod et al.\ \cite{mer87}) and proper motion (Lenham \& Franz \cite{lf61}). The Hipparcos 
data confirm these results: From two cluster members in the Hipparcos Catalogue we derive a cluster 
motion which gives a membership probability of 34\% for DL Cas. The Cepheid is therefore 
very likely a cluster member.   

V Cen (HIP 71116) is a member of NGC 5662 according to Claria et al.\ (\cite{clb}).
Figure 4a) shows the proper motions of the Cepheid and the other cluster stars. 
The agreement is excellent and the Cepheid has a high membership probability of 90\%.
We conclude that it is a member of NGC 5662. 

The membership of SZ Tau (HIP 21517) in NGC 1647 was proposed by Efremov (\cite{ev1}, 
\cite{ev2}) and confirmed by Turner (\cite{turn92}) on the basis of $UBV$ photometry
and spectroscopy. Geffert et al. (\cite{gef96}), based on proper motions
from photographic plates, denied the cluster membership of the Cepheid. We could find
five members of NGC 1647 (three certain, two possible) in the Hipparcos Catalogue.
Their mean proper motion differs significantly from the motion of the Cepheid and
rules out its membership.
Despite a rough agreement in the radial velocities of the cluster and the Cepheid 
(see the discussion in Turner \cite{turn92}), we conclude that SZ Tau is not a member of
NGC~1647.

EV Sct (HIP 91239) and Y Sct (HIP 91366) are possible members of NGC 6664. The membership
of EV Sct is well established by its radial velocity (Mermilliod et al.\ \cite{mer87}). The membership
of Y Sct is also very likely since its mean radial velocity of $\gamma = 17.8$ km/sec
(Moffett \& Barnes 1987) is in good agreement to the mean cluster velocity of $r_v = 17.8
\pm 0.2$ km/sec given by Mermilliod et al.\ (\cite{mer87}). The Hipparcos data is in 
agreement with the radial velocities,
since both Cepheids have high membership probabilities of 81.8 \%. The cluster motion
is however not very well established since the two Cepheids
are the only cluster members in the Hipparcos Catalogue. 

The situation is less clear for the Cepheids BB Sgr (HIP 92491) and GH Car (HIP 54621).
BB Sgr was proposed to be a coronal member of Col 394 by Tsarevsky et al.\ (\cite{tsar}).
Its membership was later confirmed by Turner \& Pedreros (\cite{tupe}) on the basis of $UBVRI$ 
photometry and by Gieren et al.\ (\cite{gfg}) on the basis of new calibrations of the surface 
brightness (Barnes-Evans) method. The Hipparcos data are inconclusive. Two Hipparcos members give 
a low membership probability of 2.7\% for BB Sgr, which is not completely inconsistent with the 
assumption of a cluster membership. If additional members from the TRC Catalogue are taken
into account, the membership probability of BB Sgr drops to below 0.1\%, raising serious doubts 
on its cluster membership. We note here that the relative proper motion of the Cepheid is pointing
towards the cluster, which advocates against a common origin of both.
Given also the large angular distance from the cluster
centre, we conclude that BB Sgr is unlikely to be a member of Col 394. 

GH Car is a member of Tru 18
according to Vazquez \& Feinstein (\cite{vafe}). From three stars in the Hipparcos and TRC 
catalogues (1 from Hipparcos, 2 from the TRC Catalogue), we derive a mean proper motion of
($\mu_{\alpha*}$/$\mu_\delta$) = ($-6.79 \pm 0.83$/$1.79 \pm 0.75$) mas/yr for the cluster. This 
gives a relatively low membership probability of 10\% for GH Car. However, there is  
a discrepancy in the photometric distances: From their $UBVRI$-photometry, Vazquez \& 
Feinstein (\cite{vafe}) found a cluster distance of $D = 1550$ pc. The Cepheid seems to be 
located further away, since the PL-relation of Feast \& Catchpole (\cite{feca}) gives an 
absolute magnitude of $M_V = -3.55$ and a distance of $D = 2313$ pc assuming that the 
period, mean $V$ magnitude and reddening of GH Car are given by $P =$ 5.72557~d, $<\!V\!> = 
9.17$ mag and $E(B-V) = 0.29$ mag (Vazquez \& Feinstein \cite{vafe}). GH Car is
probably not physically related to Tru 18. Radial velocities would help to confirm our 
conclusions concerning the last two Cepheids.

\begin{figure}
\resizebox{\hsize}{!}{\includegraphics{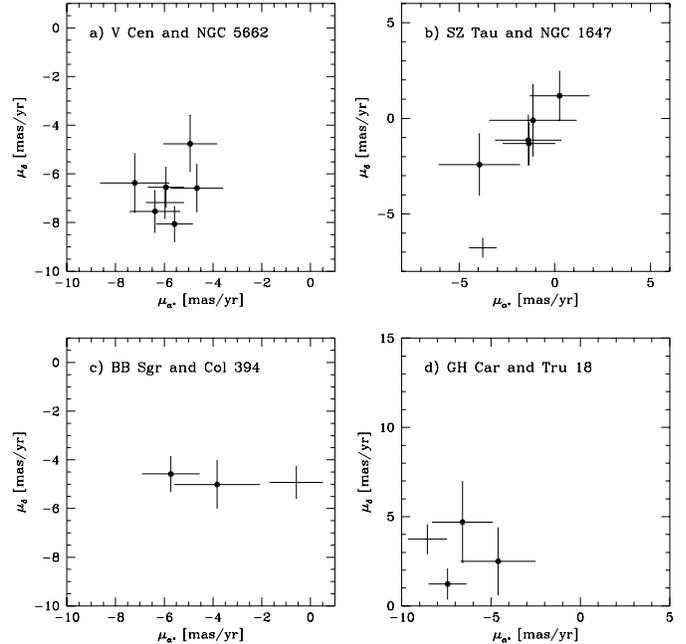}}
\caption{Proper motions of Cepheids in clusters with at least two other Hipparcos members, see Table 4.
Not shown are U~Sgr in IC~4725 and S~Nor in NGC~6087, which  
were already discussed
by Lyng{\aa} and Lindegren (\cite{ll}), and EV~Sct and Y~Sct in NGC 6664. Cluster stars are shown by 
filled circles,  
Cepheids without symbols.}
\end{figure}
\setcounter{table}{2} 
\begin{table*}[ht]
\begin{scriptsize}
{\small \caption[]{Proper motions of Cepheids and open clusters. The proper motions of the clusters
are calculated
without taking the motions of the Cepheids into account. The constants $c$ were calculated according
to eq.\ 1 and the membership probabilities in col.\ 12 were derived under the assumption
that $c$ is distributed like a chi-square distribution with two degrees of freedom.}}
\begin{tabular}{l|l|rr|rr|rr|rr|r|r|l}
\noalign{\smallskip}
\hline
\hline
\noalign{\smallskip}
 & &
\multicolumn{4}{c|}{Proper motion Cepheid} & 
\multicolumn{4}{c|}{Proper motion Cluster} & & & \\ 
\multicolumn{1}{c|}{Cepheid} &\multicolumn{1}{c|}{Cluster} & 
\multicolumn{1}{c}{$\mu_{\alpha*}$} & \multicolumn{1}{c|}{$\sigma_\mu$} &
\multicolumn{1}{c}{$\mu_\delta$} & \multicolumn{1}{c|}{$\sigma_\mu$} &
\multicolumn{1}{c}{$\mu_{\alpha*}$} & \multicolumn{1}{c|}{$\sigma_\mu$} &
\multicolumn{1}{c}{$\mu_\delta$} & \multicolumn{1}{c|}{$\sigma_\mu$} & 
\multicolumn{1}{c|}{$c$} & \multicolumn{1}{c|}{Mem.} &
\multicolumn{1}{l}{Hipparcos stars used to calculate}\\
 & & & & & & & & & & & \\[-0.18cm]
 & & \multicolumn{2}{c|}{[mas /yr]} & \multicolumn{2}{c|}{[mas /yr]} &
\multicolumn{2}{c|}{[mas /yr]} & \multicolumn{2}{c|}{[mas /yr]} & & 
\multicolumn{1}{c|}{Prob.} & \multicolumn{1}{l}{the proper motion of the cluster} \\
 & & & & & & & & & & & \\[-0.18cm]
\hline
 & & & & & & & & & & & \\[-0.13cm]
 U Sgr & IC 4725 & -4.15 & 0.94 & -6.05 & 0.70 & -1.68 & 0.87 & -6.39 & 0.62 & 3.83 &
14.7 & 90801, 90900\\[+0.1cm]
 S Nor & NGC 6087& -1.10 & 0.77 & -1.20 & 0.70 & -1.67 & 0.71 & -1.60 & 0.66 & 0.51 &
77.3 & 79891, 79907, 79973\\[+0.1cm]
 DL Cas& NGC 129 & -0.75 & 1.05 & -1.38 & 0.73 & -2.67 & 0.94 & -1.85 & 0.65 & 2.16 &
33.9 & 2354, 2382\\[+0.1cm]
 V Cen & NGC 5662& -5.97 & 0.78 & -7.18 & 0.67 & -5.60 & 0.50 & -7.33 & 0.51 & 0.21 &
90.0 & 71163, 71326, 71334, 71378, 71397,\\
 & & & & & & & & & & & & 71398\\[+0.1cm]
SZ Tau & NGC 1647& -3.76 & 0.72 & -6.77 & 0.52 & -1.37 & 0.97 & -1.02 & 0.77 & 41.6 &
0.0 & 21875, 22112, 22161, 22185, 22211\\[+0.1cm]
EV Sct & NGC 6664&  1.16 & 2.31 & -2.84 & 1.73 & -0.63 & 1.65 & -2.20 & 1.29 & 0.49 &
81.8 & 91366\\[+0.1cm]
Y Sct & NGC 6664&   -0.63 & 1.65 & -2.20 & 1.29 & 1.16 & 2.31 & -2.84 & 1.73 & 0.49 &
81.8 & 91239\\[+0.1cm] 
BB Sgr & Col 394 & -0.58 & 1.10 & -4.93 & 0.68 & -4.80 & 1.12 & -4.78 & 0.69 & 7.23 & 
2.7 & 92505, 92650\\[+0.1cm]
BB Sgr & Col 394 & -0.58 & 1.10 & -4.93 & 0.68 & -4.79 & 0.44 & -6.22 & 0.37 & 15.43 & 
0.0 & 92505, 92650, 1,6,12,22,26,27,28,33,\\
 & & & & & & & & & & & & 52,53,55,57,58,59,60,62,63,66,67,76\\[+0.1cm]
GH Car & Tru 18 &  -8.56 & 1.08 &  3.74 & 0.84 & -6.79 & 0.83 &  1.79 & 0.75 & 4.68 &
9.6 & 54668, 11, 16\\ 
\noalign{\smallskip}
\hline
\noalign{\smallskip}
\multicolumn{13}{@{}l}{Notes: Col 394: The numbers of the TRC stars in column 13 are taken from Claria et al.\ 
(\cite{clb}). Tru 18: The numbers of the two TRC stars}\\
\multicolumn{13}{@{}l}{are from Vazquez \& Feinstein (\cite{vafe}).}
\end{tabular}
\end{scriptsize}
\end{table*}

\section{Summary}

The proper motions and parallaxes of 205 open clusters were determined from 630 certain and 100 possible
members found in the Hipparcos Catalogue. A comparison of the
parallaxes with photometric distances from Loktin et al.\ (\cite{lok97}) argues for a
decrease of the photometric distance scale by $12\% \pm 6\%$, while the distance scale
of Dambis (\cite{dam99}) is in good agreement with Hipparcos.
No evidence for unaccounted small-scale correlated errors in the Hipparcos Catalogue is found
by these comparisons. It therefore seems unlikely that such errors
can explain the discrepancy between the Hipparcos parallax and photometric distance estimates 
of the Pleiades as proposed by Narayanan \& Gould (\cite{nago}). 

With the help of the Hipparcos proper motions, we can confirm the membership
of the Cepheids U Sgr in IC 4725, S Nor in NGC 6087, DL Cas in NGC 129 and V Cen in NGC 5662
and can reject the membership of SZ Tau in NGC 1647. The improved membership information
may lead to better estimates for the absolute magnitudes of Cepheids.

\begin{acknowledgements}
We are grateful to Floor van Leeuwen for calculating the correlated solutions of  
the astrometric parameters of the open clusters for us. We acknowledge the help of
Elena Glushkova and Alexander Loktin who send us their data prior to publication.
We also thank Sabine Frink and an anonymous referee for making suggestions which 
improved the paper.
Extensive use has been made of the Simbad and BDA databases. 
HB is supported by the Sonderforschungsbereich 328 Entwicklung von Galaxien.
\end{acknowledgements}

\setcounter{table}{0} 
\begin{table*}
\begin{scriptsize}
{\small \caption[]{Hipparcos stars in the fields of open clusters}}
\begin{tabular}{l|r|r|c@{}r|r||l|r|r|c@{}r|r||l|r|r|c@{}r|r}
\noalign{\smallskip}
\hline
\hline
\noalign{\smallskip}
\multicolumn{1}{c|}{Cluster} &\multicolumn{1}{c|}{HIP} & \multicolumn{1}{c|}{BDA}&
\multicolumn{2}{c|}{Mem.}&\multicolumn{1}{c||}{Prob} &
\multicolumn{1}{c|}{Cluster} &\multicolumn{1}{c|}{HIP} & \multicolumn{1}{c|}{BDA}&
\multicolumn{2}{c|}{Mem.}&\multicolumn{1}{c||}{Prob} &
\multicolumn{1}{c|}{Cluster} &\multicolumn{1}{c|}{HIP} & \multicolumn{1}{c|}{BDA}&
\multicolumn{2}{c|}{Mem.}&\multicolumn{1}{c}{Prob}\\
&\multicolumn{1}{c|}{No.}&\multicolumn{1}{c|}{No.} &\multicolumn{2}{c|}{Stat.}&
\multicolumn{1}{c||}{[\%]}& &\multicolumn{1}{c|}{No.} & \multicolumn{1}{c|}{No.}&
\multicolumn{2}{c|}{Stat.}&\multicolumn{1}{c||}{[\%]} &
&\multicolumn{1}{c|}{No.}&\multicolumn{1}{c|}{No.} & \multicolumn{2}{c|}{Stat.}&
\multicolumn{1}{c}{[\%]}\\[+0.12cm]
\hline
& & & & & & & & & & & & & & & & & \\[-0.13cm]
BER59    &      40   &        & PM & * &  84.5      & COL132   &   34954   & 4      & M &    &  16.3      & COL394   &   92505   &        & M &    &  66.7      \\
 &             139   &        & PM &   &  24.5      &  &           34964   & 10     & NM &   &   0.0      &  &           92650   & 2      & M &    &  66.7      \\
 &             172   &        & M &    &  24.5      &  &           35014   & 12     & NM &   &   1.1      & COL463   &    7878   & 36     & M &    &  88.9      \\
BER86    &  100214   & 218    & M &    &  25.8      &  &           35017   & 6      & NM &   &   0.0      &  &            7890   & 30     & NM &   &   0.0      \\
 &          100227   &        & PM &   &  25.8      &  &           35081   & 21     & NM &   &   0.0      &  &            8154   & 17     & M &    &  80.0      \\
\end{tabular}
\end{scriptsize}
\end{table*}
\clearpage

\begin{table*}
\begin{scriptsize}
\addtocounter{table}{-1}
{\small \caption[]{continued}}
\begin{tabular}{l|r|r|c@{}r|r||l|r|r|c@{}r|r||l|r|r|c@{}r|r}
\noalign{\smallskip}
\hline
\hline
\noalign{\smallskip}
\multicolumn{1}{c|}{Cluster} &\multicolumn{1}{c|}{HIP} & \multicolumn{1}{c|}{BDA}&
\multicolumn{2}{c|}{Mem.}&\multicolumn{1}{c||}{Prob} &
\multicolumn{1}{c|}{Cluster} &\multicolumn{1}{c|}{HIP} & \multicolumn{1}{c|}{BDA}&
\multicolumn{2}{c|}{Mem.}&\multicolumn{1}{c||}{Prob} &
\multicolumn{1}{c|}{Cluster} &\multicolumn{1}{c|}{HIP} & \multicolumn{1}{c|}{BDA}&
\multicolumn{2}{c|}{Mem.}&\multicolumn{1}{c}{Prob}\\
&\multicolumn{1}{c|}{No.}&\multicolumn{1}{c|}{No.} &\multicolumn{2}{c|}{Stat.}&
\multicolumn{1}{c||}{[\%]}& &\multicolumn{1}{c|}{No.} & \multicolumn{1}{c|}{No.}&
\multicolumn{2}{c|}{Stat.}&\multicolumn{1}{c||}{[\%]} &
&\multicolumn{1}{c|}{No.}&\multicolumn{1}{c|}{No.} & \multicolumn{2}{c|}{Stat.}&
\multicolumn{1}{c}{[\%]}\\[+0.12cm]
\hline
& & & & & & & & & & & & & & & & & \\[-0.13cm]
IC4665   &   87201   & 109    & NM &   &  90.7      & NGC654   &    8106   & 1      & M &    &  45.1      & NGC1647  &   22185   & 94     & M &    &  86.9      \\
 &           87202   & 108    & NM &   &   0.0      & NGC659   &    8125   & 183    & M &    &  -         &  &           22211   & 105    & PM &   &  97.3      \\
IC4725   &   90790   & 26     & NM &   &   0.0      & NGC663   &    8239   & 54     & M& *   &  61.9      & NGC1662  &   22324   & 2      & M &    &  22.9      \\
 &           90801   & 50     & M &    &  61.4      &  &            8260   & 44     & M &    &  94.3      &  &           22348   & 6      & M &    &   7.1      \\
 &           90836   & 103    & M &    &  14.9      &  &            8305   & 147    & M &    &  62.7      &  &           22351   &        & PM &   &  61.3      \\
\end{tabular}
\end{scriptsize}
\end{table*}
\clearpage

\begin{table*}
\begin{scriptsize}
\addtocounter{table}{-1}
{\small \caption[]{continued}}
\begin{tabular}{l|r|r|c@{}r|r||l|r|r|c@{}r|r||l|r|r|c@{}r|r}
\noalign{\smallskip}
\hline
\hline
\noalign{\smallskip}
\multicolumn{1}{c|}{Cluster} &\multicolumn{1}{c|}{HIP} & \multicolumn{1}{c|}{BDA}&
\multicolumn{2}{c|}{Mem.}&\multicolumn{1}{c||}{Prob} &
\multicolumn{1}{c|}{Cluster} &\multicolumn{1}{c|}{HIP} & \multicolumn{1}{c|}{BDA}&
\multicolumn{2}{c|}{Mem.}&\multicolumn{1}{c||}{Prob} &
\multicolumn{1}{c|}{Cluster} &\multicolumn{1}{c|}{HIP} & \multicolumn{1}{c|}{BDA}&
\multicolumn{2}{c|}{Mem.}&\multicolumn{1}{c}{Prob}\\
&\multicolumn{1}{c|}{No.}&\multicolumn{1}{c|}{No.} &\multicolumn{2}{c|}{Stat.}&
\multicolumn{1}{c||}{[\%]}& &\multicolumn{1}{c|}{No.} & \multicolumn{1}{c|}{No.}&
\multicolumn{2}{c|}{Stat.}&\multicolumn{1}{c||}{[\%]} &
&\multicolumn{1}{c|}{No.}&\multicolumn{1}{c|}{No.} & \multicolumn{2}{c|}{Stat.}&
\multicolumn{1}{c}{[\%]}\\[+0.12cm]
\hline
& & & & & & & & & & & & & & & & & \\[-0.13cm]
NGC2244  &   31305   & 389    & PM &   &  50.2      & NGC2439  &   37418   & 2      & M &    &  67.4      & NGC2546  &   40220   & 575    & NM &   &   0.0      \\
NGC2251  &   31365   & 1      & PM & * &  -         &  &           37433   & 3      & M &    &  50.1      &  &           40316   & 540    & NM &   &   0.0      \\
 &           31408   & 5      & NM &   &  -         & NGC2447  &   37690   &        & PM &   &  32.9      & NGC2547  &   39773   & 3      & NM &   &   0.0      \\
NGC2264  &   31939   & 50     & M &    &  86.7      &  &           37729   & 3      & M& *   &  44.3      &  &           39873   & 8      & NM &   &   0.0      \\
 &           31951   &        & M &    &  54.4      &  &           37738   & 42     & M &    &  57.3      &  &           39919   & 1      & NM &   &   0.0      \\
\end{tabular}
\end{scriptsize}
\end{table*}
\clearpage

\begin{table*}
\begin{scriptsize}
\addtocounter{table}{-1}
{\small \caption[]{continued}}
\begin{tabular}{l|r|r|c@{}r|r||l|r|r|c@{}r|r||l|r|r|c@{}r|r}
\noalign{\smallskip}
\hline
\hline
\noalign{\smallskip}
\multicolumn{1}{c|}{Cluster} &\multicolumn{1}{c|}{HIP} & \multicolumn{1}{c|}{BDA}&
\multicolumn{2}{c|}{Mem.}&\multicolumn{1}{c||}{Prob} &
\multicolumn{1}{c|}{Cluster} &\multicolumn{1}{c|}{HIP} & \multicolumn{1}{c|}{BDA}&
\multicolumn{2}{c|}{Mem.}&\multicolumn{1}{c||}{Prob} &
\multicolumn{1}{c|}{Cluster} &\multicolumn{1}{c|}{HIP} & \multicolumn{1}{c|}{BDA}&
\multicolumn{2}{c|}{Mem.}&\multicolumn{1}{c}{Prob}\\
&\multicolumn{1}{c|}{No.}&\multicolumn{1}{c|}{No.} &\multicolumn{2}{c|}{Stat.}&
\multicolumn{1}{c||}{[\%]}& &\multicolumn{1}{c|}{No.} & \multicolumn{1}{c|}{No.}&
\multicolumn{2}{c|}{Stat.}&\multicolumn{1}{c||}{[\%]} &
&\multicolumn{1}{c|}{No.}&\multicolumn{1}{c|}{No.} & \multicolumn{2}{c|}{Stat.}&
\multicolumn{1}{c}{[\%]}\\[+0.12cm]
\hline
& & & & & & & & & & & & & & & & & \\[-0.13cm]
NGC3532  &   54197   & 49     & M &    &  85.4      & NGC5749  &   72424   & 72     & M &    &  -         & NGC6475  &   87460   & 26     & M &    &  65.6      \\
 &           54237   & 401    & M &    &  98.5      &  &           72460   & 39     & NM &   &   2.9      &  &           87472   & 134    & M &    &  79.3      \\
 &           54294   & 221    & NM &   &  61.9      &  &           72519   & 7      & NM &   &   1.5      &  &           87516   &        & M &    &  97.5      \\
 &           54306   & 215    & M &    &  53.5      & NGC5822  &   73711   & 1      & M &    &  23.5      &  &           87529   & 42     & M &    &  91.5      \\
 &           54337   & 345    & M &    &  86.5      &  &           73782   & 9      & M &    &  23.5      &  &           87560   & 55     & M &    &  93.5      \\
\end{tabular}
\end{scriptsize}
\end{table*}
\clearpage

\begin{table*}
\begin{scriptsize}
\addtocounter{table}{-1}
{\small \caption[]{continued}}
\begin{tabular}{l|r|r|c@{}r|r||l|r|r|c@{}r|r||l|r|r|c@{}r|r}
\noalign{\smallskip}
\hline
\hline
\noalign{\smallskip}
\multicolumn{1}{c|}{Cluster} &\multicolumn{1}{c|}{HIP} & \multicolumn{1}{c|}{BDA}&
\multicolumn{2}{c|}{Mem.}&\multicolumn{1}{c||}{Prob} &
\multicolumn{1}{c|}{Cluster} &\multicolumn{1}{c|}{HIP} & \multicolumn{1}{c|}{BDA}&
\multicolumn{2}{c|}{Mem.}&\multicolumn{1}{c||}{Prob} &
\multicolumn{1}{c|}{Cluster} &\multicolumn{1}{c|}{HIP} & \multicolumn{1}{c|}{BDA}&
\multicolumn{2}{c|}{Mem.}&\multicolumn{1}{c}{Prob}\\
&\multicolumn{1}{c|}{No.}&\multicolumn{1}{c|}{No.} &\multicolumn{2}{c|}{Stat.}&
\multicolumn{1}{c||}{[\%]}& &\multicolumn{1}{c|}{No.} & \multicolumn{1}{c|}{No.}&
\multicolumn{2}{c|}{Stat.}&\multicolumn{1}{c||}{[\%]} &
&\multicolumn{1}{c|}{No.}&\multicolumn{1}{c|}{No.} & \multicolumn{2}{c|}{Stat.}&
\multicolumn{1}{c}{[\%]}\\[+0.12cm]
\hline
& & & & & & & & & & & & & & & & & \\[-0.13cm]
NGC6823  &   97003   & 189    & NM &   &   0.0      & NGC7160  &  107984   & 1      & M &    &  37.9      & ONC      &   26237   & 1970   & M& *   &   0.0      \\
 &           97057   & 35     & PM &   &  31.1      &  &          108052   & 4      & M &    &  80.1      &  &           26241   & 2037   & M &    &  61.8      \\
NGC6830  &   97673   & 7      & NM &   &   0.6      &  &          108073   & 2      & M &    &  84.1      &  &           26258   & 2074   & M& *   &   0.1      \\
 &           97713   & 72     & M &    &  -         &  &          108080   & 3      & M &    &   5.7      &  &           26268   & 2131   & NM &   &   0.0      \\
 &           97716   & 78     & NM &   &   0.0      & NGC7209  &  108916   & 13     & NM &   &   0.0      &  &           26314   & 2366   & M &    &  29.2      \\
\end{tabular}
\end{scriptsize}
\end{table*}
\clearpage

\begin{table*}
\begin{scriptsize}
\addtocounter{table}{-1}
{\small \caption[]{continued}}
\begin{tabular}{l|r|r|c@{}r|r||l|r|r|c@{}r|r||l|r|r|c@{}r|r}
\noalign{\smallskip}
\hline
\hline
\noalign{\smallskip}
\multicolumn{1}{c|}{Cluster} &\multicolumn{1}{c|}{HIP} & \multicolumn{1}{c|}{BDA}&
\multicolumn{2}{c|}{Mem.}&\multicolumn{1}{c||}{Prob} &
\multicolumn{1}{c|}{Cluster} &\multicolumn{1}{c|}{HIP} & \multicolumn{1}{c|}{BDA}&
\multicolumn{2}{c|}{Mem.}&\multicolumn{1}{c||}{Prob} &
\multicolumn{1}{c|}{Cluster} &\multicolumn{1}{c|}{HIP} & \multicolumn{1}{c|}{BDA}&
\multicolumn{2}{c|}{Mem.}&\multicolumn{1}{c}{Prob}\\
&\multicolumn{1}{c|}{No.}&\multicolumn{1}{c|}{No.} &\multicolumn{2}{c|}{Stat.}&
\multicolumn{1}{c||}{[\%]}& &\multicolumn{1}{c|}{No.} & \multicolumn{1}{c|}{No.}&
\multicolumn{2}{c|}{Stat.}&\multicolumn{1}{c||}{[\%]} &
&\multicolumn{1}{c|}{No.}&\multicolumn{1}{c|}{No.} & \multicolumn{2}{c|}{Stat.}&
\multicolumn{1}{c}{[\%]}\\[+0.12cm]
\hline
& & & & & & & & & & & & & & & & & \\[-0.13cm]
TRU10    &   43240   & 6      & M &    &  95.6      & TRU37    &  106716   & 710    & NM &   &   0.0      & TRU37    &  107353   & 1331   & NM &   &   0.0      \\
 &           43326   &        & PM &   &  38.3      &  &          106774   & 452    & PM & * &   0.0      &  &          107500   & 225    & NM &   &   0.0      \\
TRU14    &   52558   & 101    & M &    &  -         &  &          106830   & 1293   & NM &   &   8.7      &  &          107598   & 1588   & NM &   &   0.8      \\
TRU15    &   52562   & 16     & PM &   &  -         &  &          106843   & 171    & M& *   &  34.3      & UPG1     &   61383   & 5      & NM &   &  -         \\
TRU16    &   52558   & 180    & NM &   &  -         &  &          106884   & 462    & M& *   &  13.7      &  &           61424   & 1      & NM &   &  -         \\
\end{tabular}
\end{scriptsize}
\end{table*}
\clearpage

\begin{small}
\noindent
Notes to Table 1:\\[+0.2cm]
{\bf Ber 59:} The star cluster Ber 59 is embedded in the gas and dust complex Sharpless 171/NGC 7822. It is 
 called NGC 7822 by some authors, but we prefer Lyng{\aa}'s notation.\\
{\bf Col 121:} HIP 33316, HIP 33846, HIP 34444, HIP 34489 and HIP 34954 are non-members due to their 
 large angular distance from the cluster centre. On the basis of their proper motions, they may be part of an
 unbound association of stars which shares the same kinematics as Col 121.\\
{\bf Col 132:} It is not clear whether this is a true star cluster, see Claria (\cite{cla77a}) and Eggen (\cite{egg83}).
  Members and non-members are taken from Baumgardt (\cite{baumg}).\\ 
{\bf Col 135:} There is some discussion in the literature about the existence of this cluster. The Hipparcos data 
 strongly supports its reality, see Baumgardt (\cite{baumg}). Members and non-members are taken from this work.\\
{\bf Col 140:} The membership of HIP 35905 in this cluster is well established (see for example Williams \cite{will}
 and Jenkner \& Maitzen \cite{jenk}). Nevertheless its proper motion deviates strongly from the other members. We
 therefore did not use it to determine the mean cluster motion.\\
{\bf Col 197:} There is an apparent clustering of possible members in the TRC Catalogue near 
 ($\mu_{\alpha*}$/$\mu_{\delta}$) = (-3/9) mas/yr. Since the proper motion of HIP 42908 is
 ($\mu_{\alpha*}$/$\mu_{\delta}$) = ($-6.85 \pm 0.82$/$ 3.22 \pm 0.87$) mas/yr it cannot be a member.\\
{\bf Col 359:} The most recent study of this cluster was made by Ruci\'nski (\cite{ruc}), who found 9 probable members and 
 a distance modulus of $V-M_V$ = 8.2 mag. Seven of the possible members are included in the Hipparcos Catalogue, but only two share
  a common proper motion. Three other Hipparcos stars may be additional cluster members, but the reality of the cluster seems doubtful.\\
{\bf Col 394:} Concerning the membership of BB Sgr (HIP 92491) see the discussion in the text.\\
{\bf HM 1:} The cluster Havlen-Moffat 1.\\
{\bf Hog 14:} It is not clear whether this cluster exists, see Moffat \& Vogt (\cite{mv73}).
 According to these authors, HIP 60863 could be a member if the cluster is real.\\
{\bf IC 1848:} The Hipparcos proper motion of HIP 13308 is incompatible with the assumption of
 cluster membership. The star is nevertheless almost certainly a cluster member on the basis of its
  radial velocity (Liu et al.\  
  \cite{liu91}) and UBV photometry. In addition its TRC proper motion is close to that of the other members.
  We therefore consider HIP 13308 to be a member but have not used it to derive the mean cluster motion.\\
{\bf IC 2395:} The Hipparcos stars show no evidence for a clustering. There is however a
 clustering of the possible members in the TRC Catalogue near ($\mu_{\alpha*}$/$\mu_{\delta}$) = (-4/1) mas/yr.
 Keeping this in mind, we made the classification of the Hipparcos stars.\\
{\bf IC 2944:} HIP 56134, HIP 56890, HIP 56961 and HIP 57057 are not considered to be cluster members due to their
 large angular distance from the cluster centre. On the basis of their proper motions, they may be part of an
 unbound association of stars which shares the kinematics of IC 2944.\\ 
{\bf IC 4665:} HIP 86944 and HIP 87184 are almost certainly members on the basis of their radial velocity, ground-based
 proper motion (Prosser \cite{pros}) and photometry. In addition their TRC proper motions are consistent with cluster 
  membership. The most likely explanation is that the Hipparcos proper motions are disturbed by binary effects. They
   were not used to determine the mean cluster motion.\\ 
{\bf Mel 227 + NGC 1252:} These are not open clusters, see the discussion in Baumgardt (\cite{baumg}).\\
{\bf NGC 654:} HIP 7936 is regarded as a possible member since it has a 92\% membership probability
in the proper motion study of Stone (\cite{stone}).\\
{\bf NGC 1647:} Concerning the membership of SZ Tau (HIP 21517) see the discussion in the text.\\
{\bf NGC 1907:} Glushkova \& Rastorguev (\cite{gl91}) found several
 giants with $V$-magnitudes between 11.9 and 12.5 mag to be cluster members. Except
 for HIP 25599, all Hipparcos stars are therefore too bright to be cluster
 members. HIP 25599 may be a cluster member in the blue straggler phase.\\
{\bf NGC 2129:} The existence of this cluster is doubtful, see Pe\~na \& Peniche (\cite{pena}).
 However, the radial velocities (Liu et al. \cite{liu91}) and the proper motions of the two Hipparcos
 stars agree with each other. This may be a hint for the reality of the cluster.\\
{\bf NGC 2251:} The membership of HIP 31365 in this cluster is uncertain, see Maitzen et al. (\cite{mai81}).\\
{\bf NGC 2362:} HIP 35461 was considered to be a cluster member by Perry (\cite{perry}) on the basis of its $UBV$ photometry. 
Its Hipparcos proper motion differs considerably from the other stars.
Since the star is known to be double, the Hipparcos
 proper motion may be in error and we have not used it to derive the mean cluster parameters.\\
{\bf NGC 2422:} HIP 36736 and HIP 36773 are not used to determine the mean cluster parameters
due to their discrepant proper motions. They are possible members on 
 the basis of their radial velocities, photometry and proper motions in the TRC Catalogue.\\
{\bf NGC 2437:} The proper motion of HIP 37480 is not well determined by Hipparcos. The star seems to be a non-member on the basis of 
  its proper motion in the TRC Catalogue.\\
{\bf NGC 2451:} There is growing evidence for the presence of two star clusters in the field of NGC 2451, see 
 R\"oser \& Bastian (\cite{roeba}), Platais et al. (\cite{pla96}) and Baumgardt (\cite{baumg}). The 
 members and possible members of NGC 2451A are taken from Baumgardt (\cite{baumg}).
HIP 37450 and HIP 37838 have somewhat discrepant proper motions. Their parallaxes are however close to the cluster 
mean. Furthermore the large proper motion of NGC 2451A clearly separates the members from the field stars. The two 
stars are therefore considered to be members.\\ 
{\bf NGC 2467:} This is not a true star cluster, see Loden (\cite{lod1}, \cite{lod2}) and Feinstein \& V\'azquez 
(\cite{feva}).\\
{\bf NGC 2483:} This is not a true star cluster, see Fitzgerald \& Moffat (\cite{fitz}) and Havlen (\cite{havl}).\\
{\bf NGC 2527:} HIP 39574 and 39602 are non-members according to Lindoff (\cite{lin73}). Their
 proper motions are nevertheless compatible with cluster membership.\\
{\bf NGC 2579:} Lindoff (\cite{lind}) found 9 possible members but had some doubt concerning the reality of the cluster. We
 could find 3 possible members in the TRC Catalogue. Their proper motions agree with each other, adding some evidence
 for the reality of the cluster. HIP 41024 is a possible member according to Lindoff (\cite{lind}), and furthermore its proper 
motion is in agreement with the two other TRC stars. It may therefore be a member.\\
{\bf NGC 2669:} HIP 43083 is not used to determine the cluster motion. The star is a probable member based on its
photometry and its proper motion in the TRC Catalogue. The Hipparcos proper motion however differs significantly 
from the mean of other members in the TRC Catalogue.
The star may be a binary.\\
{\bf NGC 3114:} HIP 49245 has a radial velocity in good agreement with that of other
cluster members in the radial velocity study of Amieux \& Burnage (\cite{ambur}). Furthermore its 
photometry is in good agreement with the assumption of a cluster membership. We therefore regard it as
a cluster member.\\
{\bf NGC 3324:} The membership of HIP 52004 in this cluster is unclear. It is a foreground star according to Claria 
 (\cite{cla77b}), but Forte (\cite{forte}) considers it to be a member on the basis of its radial velocity and
 due to an HII shell which surrounds the star and is also connected with the cluster.\\
{\bf NGC 3496:} This is probably not a true star cluster, see Sher (\cite{sher}) and Balona \& Laney (\cite{bala}). 
 Even if there is a cluster in this field, HIP 53743 is probably not a member (Sher \cite{sher}).\\
{\bf NGC 3572:} The membership of HIP 54606 is uncertain. Schmidt-Kaler (\cite{schmk}) considers the star to be a cluster
 member, while Moffat \& Vogt (\cite{mv75}) do not.\\
{\bf NGC 4337:} The photometry of Moffat \& Vogt (\cite{mv73}) casts doubts on the reality of this cluster.\\
{\bf NGC 4755:} HIP 62931 has a 54\% membership probability in the proper motion study of King (\cite{kin81}). In addition
its radial velocity, photometry and central position in the cluster strongly support its membership. We therefore regard
it as a cluster member.\\
{\bf NGC 5823:} The membership of both stars is uncertain since they are not well observed. 
Their photometry is 
 compatible with cluster membership.\\
{\bf NGC 6169:} This is not an open cluster, see Moffat \& Vogt (\cite{mv73}).\\
{\bf NGC 6618:} M17. The membership of HIP 89804 is somewhat uncertain. It is a member according to Hobbs (\cite{hobbs}).\\
{\bf NGC 6755:} The membership of both stars is somewhat uncertain since the cluster is not well observed. Svolopoulos 
 (\cite{svo}) considers HIP 93932 to be a member on the basis of its spectral type and photometry.\\
{\bf NGC 6834:} There is no consensus in the literature concerning the membership of HIP 97764 and HIP 97806. Our classification
  follows the results of Turner (\cite{turn76}).\\
{\bf NGC 6866:} The proper motions of both stars deviate considerably from each other. However, the  possible members in the TRC 
 Catalogue show a clustering near ($\mu_{\alpha*}$/$ \mu_{\delta}$) = (-2/-5) mas/yr and the proper 
 motions of both stars are compatible with this value. We therefore consider them to be cluster members.\\
{\bf NGC 6883:} The membership of HIP 99437 is very uncertain because the cluster is not well observed.\\
{\bf NGC 7092:} The proper motion difference between HIP 106346 and the other cluster stars is too large to be explained
 by a statistical fluctuation. HIP 106346 was therefore not used to determine the mean cluster motion.\\ 
{\bf NGC 7686:} This is probably not a real star cluster, see Johnson (\cite{john}) and Becker (\cite{beck}).\\
{\bf ONC:} The Orion Nebula Cluster. The following three stars were not used to determine the mean cluster motion:
 HIP 26237, 26258 and 26442. They have proper motions close to the cluster mean in the proper motion study of McNamara
 et al.\ (\cite{mcnm}), but their Hipparcos proper motions differ significantly. The most likely explanation is that 
 the Hipparcos proper motion is influenced by binary effects. The proper motions of HIP 26000 and HIP 26199 
are very close to the cluster mean in the study of McNamara et al.\ (\cite{mcnm}). They are therefore considered
to be members.\\
{\bf ROS 5:} It is unclear whether this is a true star cluster, see Lee \& Perry (\cite{lee71}) and Baumgardt (\cite{baumg}). The 
possible members are taken from Baumgardt (\cite{baumg}).\\
{\bf RUP 98:} On the basis of their proper motions HIP 58411 and HIP 58432 exclude each other as members. The mean proper motion of
 the members in the TRC Catalogue strongly supports the membership of HIP 58432.  We therefore consider HIP 58411 
 to be a field star.\\
{\bf TRU 18:} Concerning the membership of GH Car (HIP 54621) see the discussion in the text.\\
{\bf TRU 24:} The existence of this cluster is doubtful, see the discussion in Perry et al.\ (\cite{perry2}).\\
{\bf TRU 33:} Two cluster members in the TRC Catalogue have proper motions close to 
($\mu_{\alpha*}$/$ \mu_{\delta}$) = (-10.5/-2.5) mas/yr.
 Therefore, on the basis of its proper motion, HIP 90230 cannot be a cluster member.\\
{\bf UPG 1:} This is not a star cluster, see Baumgardt (\cite{baumg}).\\
\end{small}
    
\clearpage
\begin{table*}
{\caption[]{Mean astrometric parameters of open clusters. The Table is
organised as follows: After the cluster name in column 1, we give the total number of members found in Hipparcos and the number of stars used to calculate
the astrometric solution in column 2. Columns 3 to 5 give the mean parallaxes and proper motions together with their errors. Note that proper motions in right ascension
are multiplied by the cosine of the declination. Column 6 gives the total number of
abscissae and the number of rejected abscissae. Column 7 gives the unit weight standard deviation for the cluster. Columns 8 and 9 finally give the correlation
coefficients between the parallax and the proper motions and among the proper motion components.}}
\begin{tabular}{l|r|r|r|r|r|r|r|r}
\noalign{\smallskip}
\hline
\hline
\noalign{\smallskip}
\multicolumn{1}{c|}{Cluster}&
\multicolumn{1}{c|}{$N_{Tot}$}&
\multicolumn{1}{c|}{$\pi$}&\multicolumn{1}{c|}{$\mu_{\alpha*}$} & \multicolumn{1}{c|}{$\mu_\delta$}&
\multicolumn{1}{c|}{$N_{Abs}$}&\multicolumn{1}{c|}{uwsd} & \multicolumn{1}{c|}{$\rho^\pi_{\mu \alpha*}$}
& \multicolumn{1}{c}{$\rho^{\mu\alpha*}_{\mu \delta}$}\\
 & \multicolumn{1}{c|}{$N_{Use}$} & 
\multicolumn{1}{c|}{$\sigma_\pi$}&\multicolumn{1}{c|}{$\sigma_\mu$} & \multicolumn{1}{c|}{$\sigma_\mu$}
&\multicolumn{1}{c|}{$N_{Rej}$} & & \multicolumn{1}{c|}{$\rho^\pi_{\mu \delta}$}&\\
 & & 
\multicolumn{1}{c|}{[mas]} & \multicolumn{1}{c|}{[mas/yr]} &  \multicolumn{1}{c|}{[mas/yr]} & & & &\\[+0.12cm]
\hline
 & & & & & & & & \\[-0.13cm]
BER 59    &  3   &      1.36   &  -2.11   &  -1.20   &  186   &    1.07    &  -0.03   &  -0.06  \\ 
 &   2   &      1.03   &   0.81   &   0.75   &    0   &    &   0.14   & \\[+0.07cm]
BER 86    &  2   &      1.73   &  -3.80   &  -4.11   &  167   &    1.00    &  -0.04   &  -0.06  \\ 
 &   2   &      0.65   &   0.54   &   0.57   &    0   &    &   0.04   & \\
\end{tabular}
\end{table*}

\clearpage

\begin{table*}
\addtocounter{table}{-1}
{\caption[]{continued}}
\begin{tabular}{l|r|r|r|r|r|r|r|r}
\noalign{\smallskip}
\hline
\hline
\noalign{\smallskip}
\multicolumn{1}{c|}{Cluster}&
\multicolumn{1}{c|}{$N_{Tot}$}&
\multicolumn{1}{c|}{$\pi$}&\multicolumn{1}{c|}{$\mu_{\alpha*}$} & \multicolumn{1}{c|}{$\mu_\delta$}&
\multicolumn{1}{c|}{$N_{Abs}$}&\multicolumn{1}{c|}{uwsd} & \multicolumn{1}{c|}{$\rho^\pi_{\mu \alpha*}$}
& \multicolumn{1}{c}{$\rho^{\mu\alpha*}_{\mu \delta}$}\\
 & \multicolumn{1}{c|}{$N_{Use}$} & 
\multicolumn{1}{c|}{$\sigma_\pi$}&\multicolumn{1}{c|}{$\sigma_\mu$} & \multicolumn{1}{c|}{$\sigma_\mu$}
&\multicolumn{1}{c|}{$N_{Rej}$} & & \multicolumn{1}{c|}{$\rho^\pi_{\mu \delta}$}&\\
 & & 
\multicolumn{1}{c|}{[mas]} & \multicolumn{1}{c|}{[mas/yr]} &  \multicolumn{1}{c|}{[mas/yr]} & & & &\\[+0.12cm]
\hline
 & & & & & & & & \\[-0.13cm]
IC 1805   &  6   &      1.42   &  -1.10   &  -1.92   &  361   &    1.10    &   0.06   &  -0.29  \\ 
 &   5   &      0.85   &   0.78   &   0.69   &    3   &    &   0.08   & \\[+0.07cm]
IC 1848   &  3   &     -0.26   &  -5.15   &   0.35   &  &          &          -0.02   &  -0.16  \\ 
 &   1   &      1.28   &   1.32   &   1.38   &  &          &   0.10   & \\
\end{tabular}
\end{table*}

\clearpage

\begin{table*}
\addtocounter{table}{-1}
{\caption[]{continued}}
\begin{tabular}{l|r|r|r|r|r|r|r|r}
\noalign{\smallskip}
\hline
\hline
\noalign{\smallskip}
\multicolumn{1}{c|}{Cluster}&
\multicolumn{1}{c|}{$N_{Tot}$}&
\multicolumn{1}{c|}{$\pi$}&\multicolumn{1}{c|}{$\mu_{\alpha*}$} & \multicolumn{1}{c|}{$\mu_\delta$}&
\multicolumn{1}{c|}{$N_{Abs}$}&\multicolumn{1}{c|}{uwsd} & \multicolumn{1}{c|}{$\rho^\pi_{\mu \alpha*}$}
& \multicolumn{1}{c}{$\rho^{\mu\alpha*}_{\mu \delta}$}\\
 & \multicolumn{1}{c|}{$N_{Use}$} & 
\multicolumn{1}{c|}{$\sigma_\pi$}&\multicolumn{1}{c|}{$\sigma_\mu$} & \multicolumn{1}{c|}{$\sigma_\mu$}
&\multicolumn{1}{c|}{$N_{Rej}$} & & \multicolumn{1}{c|}{$\rho^\pi_{\mu \delta}$}&\\
 & & 
\multicolumn{1}{c|}{[mas]} & \multicolumn{1}{c|}{[mas/yr]} &  \multicolumn{1}{c|}{[mas/yr]} & & & &\\[+0.12cm]
\hline
 & & & & & & & & \\[-0.13cm]
NGC 869   &  4   &      0.66   &  -0.92   &  -1.66   &  283   &    0.93    &   0.06   &  -0.24  \\ 
 &   4   &      0.45   &   0.37   &   0.34   &    2   &    &   0.18   & \\[+0.07cm]
NGC 884   &  4   &      0.80   &  -1.14   &  -1.56   &  277   &    0.94    &   0.03   &  -0.24  \\ 
 &   4   &      0.50   &   0.43   &   0.38   &    0   &    &   0.18   & \\
\end{tabular}
\end{table*}

\clearpage

\begin{table*}
\addtocounter{table}{-1}
{\caption[]{continued}}
\begin{tabular}{l|r|r|r|r|r|r|r|r}
\noalign{\smallskip}
\hline
\hline
\noalign{\smallskip}
\multicolumn{1}{c|}{Cluster}&
\multicolumn{1}{c|}{$N_{Tot}$}&
\multicolumn{1}{c|}{$\pi$}&\multicolumn{1}{c|}{$\mu_{\alpha*}$} & \multicolumn{1}{c|}{$\mu_\delta$}&
\multicolumn{1}{c|}{$N_{Abs}$}&\multicolumn{1}{c|}{uwsd} & \multicolumn{1}{c|}{$\rho^\pi_{\mu \alpha*}$}
& \multicolumn{1}{c}{$\rho^{\mu\alpha*}_{\mu \delta}$}\\
 & \multicolumn{1}{c|}{$N_{Use}$} & 
\multicolumn{1}{c|}{$\sigma_\pi$}&\multicolumn{1}{c|}{$\sigma_\mu$} & \multicolumn{1}{c|}{$\sigma_\mu$}
&\multicolumn{1}{c|}{$N_{Rej}$} & & \multicolumn{1}{c|}{$\rho^\pi_{\mu \delta}$}&\\
 & & 
\multicolumn{1}{c|}{[mas]} & \multicolumn{1}{c|}{[mas/yr]} &  \multicolumn{1}{c|}{[mas/yr]} & & & &\\[+0.12cm]
\hline
 & & & & & & & & \\[-0.13cm]
NGC 2244  &  5   &      1.07   &  -0.65   &   0.67   &  166   &    0.95    &  -0.01   &  -0.18  \\ 
 &   5   &      0.57   &   0.47   &   0.39   &    1   &    &  -0.08   & \\[+0.07cm]
NGC 2264  &  6   &      2.94   &  -0.52   &  -4.47   &  213   &    0.91    &  -0.10   &   0.30  \\ 
 &   5   &      0.62   &   0.65   &   0.44   &    1   &    &  -0.06   & \\
\end{tabular}
\end{table*}

\clearpage

\begin{table*}
\addtocounter{table}{-1}
{\caption[]{continued}}
\begin{tabular}{l|r|r|r|r|r|r|r|r}
\noalign{\smallskip}
\hline
\hline
\noalign{\smallskip}
\multicolumn{1}{c|}{Cluster}&
\multicolumn{1}{c|}{$N_{Tot}$}&
\multicolumn{1}{c|}{$\pi$}&\multicolumn{1}{c|}{$\mu_{\alpha*}$} & \multicolumn{1}{c|}{$\mu_\delta$}&
\multicolumn{1}{c|}{$N_{Abs}$}&\multicolumn{1}{c|}{uwsd} & \multicolumn{1}{c|}{$\rho^\pi_{\mu \alpha*}$}
& \multicolumn{1}{c}{$\rho^{\mu\alpha*}_{\mu \delta}$}\\
 & \multicolumn{1}{c|}{$N_{Use}$} & 
\multicolumn{1}{c|}{$\sigma_\pi$}&\multicolumn{1}{c|}{$\sigma_\mu$} & \multicolumn{1}{c|}{$\sigma_\mu$}
&\multicolumn{1}{c|}{$N_{Rej}$} & & \multicolumn{1}{c|}{$\rho^\pi_{\mu \delta}$}&\\
 & & 
\multicolumn{1}{c|}{[mas]} & \multicolumn{1}{c|}{[mas/yr]} &  \multicolumn{1}{c|}{[mas/yr]} & & & &\\[+0.12cm]
\hline
 & & & & & & & & \\[-0.13cm]
NGC 2546  &  3   &      1.05   &  -2.92   &   4.71   &  175   &    0.90    &   0.03   &  -0.05  \\ 
 &   2   &      0.56   &   0.42   &   0.49   &    2   &    &   0.10   & \\[+0.07cm]
NGC 2547  &  8   &      2.27   &  -9.56   &   4.42   &  598   &    0.88    &   0.10   &   0.03  \\ 
 &   8   &      0.29   &   0.32   &   0.24   &    1   &    &   0.13   & \\
\end{tabular}
\end{table*}

\clearpage

\begin{table*}
\addtocounter{table}{-1}
{\caption[]{continued}}
\begin{tabular}{l|r|r|r|r|r|r|r|r}
\noalign{\smallskip}
\hline
\hline
\noalign{\smallskip}
\multicolumn{1}{c|}{Cluster}&
\multicolumn{1}{c|}{$N_{Tot}$}&
\multicolumn{1}{c|}{$\pi$}&\multicolumn{1}{c|}{$\mu_{\alpha*}$} & \multicolumn{1}{c|}{$\mu_\delta$}&
\multicolumn{1}{c|}{$N_{Abs}$}&\multicolumn{1}{c|}{uwsd} & \multicolumn{1}{c|}{$\rho^\pi_{\mu \alpha*}$}
& \multicolumn{1}{c}{$\rho^{\mu\alpha*}_{\mu \delta}$}\\
 & \multicolumn{1}{c|}{$N_{Use}$} & 
\multicolumn{1}{c|}{$\sigma_\pi$}&\multicolumn{1}{c|}{$\sigma_\mu$} & \multicolumn{1}{c|}{$\sigma_\mu$}
&\multicolumn{1}{c|}{$N_{Rej}$} & & \multicolumn{1}{c|}{$\rho^\pi_{\mu \delta}$}&\\
 & & 
\multicolumn{1}{c|}{[mas]} & \multicolumn{1}{c|}{[mas/yr]} &  \multicolumn{1}{c|}{[mas/yr]} & & & &\\[+0.12cm]
\hline
 & & & & & & & & \\[-0.13cm]
NGC 5460  &  2   &      0.62   &  -7.04   &  -3.23   &  121   &    0.94    &  -0.14   &  -0.10  \\ 
 &   2   &      0.76   &   0.72   &   0.64   &    0   &    &   0.04   & \\[+0.07cm]
NGC 5606  &  3   &     -1.27   &  -5.19   &  -4.60   &  184   &    1.00    &  -0.02   &   0.48  \\ 
 &   3   &      0.94   &   1.06   &   0.80   &    0   &    &  -0.01   & \\
\end{tabular}
\end{table*}

\clearpage

\begin{table*}
\addtocounter{table}{-1}
{\caption[]{continued}}
\begin{tabular}{l|r|r|r|r|r|r|r|r}
\noalign{\smallskip}
\hline
\hline
\noalign{\smallskip}
\multicolumn{1}{c|}{Cluster}&
\multicolumn{1}{c|}{$N_{Tot}$}&
\multicolumn{1}{c|}{$\pi$}&\multicolumn{1}{c|}{$\mu_{\alpha*}$} & \multicolumn{1}{c|}{$\mu_\delta$}&
\multicolumn{1}{c|}{$N_{Abs}$}&\multicolumn{1}{c|}{uwsd} & \multicolumn{1}{c|}{$\rho^\pi_{\mu \alpha*}$}
& \multicolumn{1}{c}{$\rho^{\mu\alpha*}_{\mu \delta}$}\\
 & \multicolumn{1}{c|}{$N_{Use}$} & 
\multicolumn{1}{c|}{$\sigma_\pi$}&\multicolumn{1}{c|}{$\sigma_\mu$} & \multicolumn{1}{c|}{$\sigma_\mu$}
&\multicolumn{1}{c|}{$N_{Rej}$} & & \multicolumn{1}{c|}{$\rho^\pi_{\mu \delta}$}&\\
 & & 
\multicolumn{1}{c|}{[mas]} & \multicolumn{1}{c|}{[mas/yr]} &  \multicolumn{1}{c|}{[mas/yr]} & & & &\\[+0.12cm]
\hline
 & & & & & & & & \\[-0.13cm]
NGC 6475  & 18   &      3.66   &   2.61   &  -4.80   &  531   &    0.77    &  -0.12   &  -0.13  \\ 
 &  16   &      0.30   &   0.33   &   0.22   &    0   &    &   0.02   & \\[+0.07cm]
NGC 6494  &  3   &     -1.37   &  -1.30   &  -0.67   &  157   &    1.14    &  -0.17   &   0.02  \\ 
 &   3   &      1.32   &   1.42   &   0.80   &    0   &    &   0.12   & \\
\end{tabular}
\end{table*}

\clearpage

\begin{table*}
\addtocounter{table}{-1}
{\caption[]{continued}}
\begin{tabular}{l|r|r|r|r|r|r|r|r}
\noalign{\smallskip}
\hline
\hline
\noalign{\smallskip}
\multicolumn{1}{c|}{Cluster}&
\multicolumn{1}{c|}{$N_{Tot}$}&
\multicolumn{1}{c|}{$\pi$}&\multicolumn{1}{c|}{$\mu_{\alpha*}$} & \multicolumn{1}{c|}{$\mu_\delta$}&
\multicolumn{1}{c|}{$N_{Abs}$}&\multicolumn{1}{c|}{uwsd} & \multicolumn{1}{c|}{$\rho^\pi_{\mu \alpha*}$}
& \multicolumn{1}{c}{$\rho^{\mu\alpha*}_{\mu \delta}$}\\
 & \multicolumn{1}{c|}{$N_{Use}$} & 
\multicolumn{1}{c|}{$\sigma_\pi$}&\multicolumn{1}{c|}{$\sigma_\mu$} & \multicolumn{1}{c|}{$\sigma_\mu$}
&\multicolumn{1}{c|}{$N_{Rej}$} & & \multicolumn{1}{c|}{$\rho^\pi_{\mu \delta}$}&\\
 & & 
\multicolumn{1}{c|}{[mas]} & \multicolumn{1}{c|}{[mas/yr]} &  \multicolumn{1}{c|}{[mas/yr]} & & & &\\[+0.12cm]
\hline
 & & & & & & & & \\[-0.13cm]
NGC 6940  &  3   &     -0.47   &  -2.23   &  -9.70   &  130   &    1.13    &  -0.08   &   0.13  \\ 
 &   2   &      1.20   &   0.88   &   0.89   &    0   &    &  -0.02   & \\[+0.07cm]
NGC 7039  &  5   &      1.55   &  -0.48   &  -2.99   &  163   &    0.89    &  -0.02   &   0.08  \\ 
 &   2   &      0.51   &   0.50   &   0.47   &    0   &    &   0.04   & \\
\end{tabular}
\end{table*}

\clearpage

\begin{table*}
\addtocounter{table}{-1}
{\caption[]{continued}}
\begin{tabular}{l|r|r|r|r|r|r|r|r}
\noalign{\smallskip}
\hline
\hline
\noalign{\smallskip}
\multicolumn{1}{c|}{Cluster}&
\multicolumn{1}{c|}{$N_{Tot}$}&
\multicolumn{1}{c|}{$\pi$}&\multicolumn{1}{c|}{$\mu_{\alpha*}$} & \multicolumn{1}{c|}{$\mu_\delta$}&
\multicolumn{1}{c|}{$N_{Abs}$}&\multicolumn{1}{c|}{uwsd} & \multicolumn{1}{c|}{$\rho^\pi_{\mu \alpha*}$}
& \multicolumn{1}{c}{$\rho^{\mu\alpha*}_{\mu \delta}$}\\
 & \multicolumn{1}{c|}{$N_{Use}$} & 
\multicolumn{1}{c|}{$\sigma_\pi$}&\multicolumn{1}{c|}{$\sigma_\mu$} & \multicolumn{1}{c|}{$\sigma_\mu$}
&\multicolumn{1}{c|}{$N_{Rej}$} & & \multicolumn{1}{c|}{$\rho^\pi_{\mu \delta}$}&\\
 & & 
\multicolumn{1}{c|}{[mas]} & \multicolumn{1}{c|}{[mas/yr]} &  \multicolumn{1}{c|}{[mas/yr]} & & & &\\[+0.12cm]
\hline
 & & & & & & & & \\[-0.13cm]
TRU 15    &  1   &      1.49   &  -7.71   &   3.26   &  &          &           0.10   &   0.07  \\ 
 &   1   &      0.90   &   1.03   &   0.78   &  &          &   0.02   & \\[+0.07cm]
TRU 17    &  1   &     -1.62   &  -2.59   &   2.88   &  &          &          -0.12   &   0.18  \\ 
 &   1   &      1.83   &   1.97   &   1.61   &  &          &   0.13   & \\
\end{tabular}
\end{table*}

\end{document}